# Coherently driving a single quantum two-level system with dichromatic laser pulses

Yu-Ming He[*], Hui Wang[*], Can Wang[*], Ming-Cheng Chen, Xing Ding, Jian Qin,
Zhao-Chen Duan, Si Chen, Jin-Peng Li, Run-Ze Liu, Christian Schneider,
Mete Atatüre, Sven Höfling, Chao-Yang Lu[#], Jian-Wei Pan[$]

[1] *Hefei National Laboratory for Physical Sciences at Microscale and Department of Modern Physics, University of Science and Technology of China, Hefei, Anhui, 230026, China*

[2] *CAS Centre for Excellence and Synergetic Innovation Centre in Quantum Information and Quantum Physics, University of Science and Technology of China, Hefei, Anhui 230026, China*

[3] *Cavendish Laboratory, University of Cambridge, JJ Thomson Ave., Cambridge CB3 0HE, United Kingdom*

[4] *Technische Physik, Physikalisches Instität and Wilhelm Conrad Röntgen-Center for Complex Material Systems, Universitat Würzburg, Am Hubland, D-97074 Würzburg, Germany*

[5] *SUPA, School of Physics and Astronomy, University of St. Andrews, St. Andrews, KY16 9SS, United Kingdom*

*\* These authors contributed equally to this work*

*Corresponding author # cylu@ustc.edu.cn, $ pan@ustc.edu.cn,*

**Efficient excitation of a single two-level system usually requires that the driving field is at the same frequency as the atomic transition. However, the scattered laser light in solid-state implementations can dominate over the single photons, imposing an outstanding challenge to perfect single-photon sources. Here, we propose a background-free method using a phase-locked dichromatic electromagnetic field with no spectral overlap with the optical transition for a coherent control of a two-level system, and we demonstrate this method experimentally with a single quantum dot embedded in a micropillar. Single photons generated by $\pi$ excitation show a purity of 0.988(1) and indistinguishability of 0.962(6). Further, the phase-coherent nature of the two-color excitation is captured by the resonance-fluorescence intensity dependence on the relative phase between the two pulses.**

**Our two-color excitation method adds a useful toolbox to the study of atom-photon interaction, and the generation of spectrally isolated indistinguishable single photons.**

The coherent control of a single two-level quantum system using electromagnetic field, by tuning the frequency, amplitude, and envelope of the field, is the elemental tool of quantum optics: Resonant atom-photon interaction underpins interesting phenomena such as Rabi oscillation[1], Ramsey interference[2], Autler-Townes splitting[3] and Mollow triplet[4]. Meanwhile, it has also become ubiquitous in quantum information technologies[5], used extensively for the initialization, manipulation and measurement of quantum bits in various physical systems including trapped ions[6], quantum dots[7-9], defects in solids[10], and superconducting quantum circuits[11].

The resonant field can rotate the quantum state arbitrarily on the XZ plane of the Bloch sphere. Specially, a π pulse excitation changes the quantum state between the two antipodes on the Bloch sphere[12,13]. A typical application is resonance fluorescence single-photon sources[9,14-20] from quantum dots, which have shown near-unity purity and indistinguishability, and high extraction efficiency[21-23]. Because the pumping laser spectrally overlaps with the single photons, polarization filtering was usually used to suppress the laser background which, however, reduced the system efficiency of the single-photon source[21-23] by at least 50%. The quest of large-scale boson sampling[24] and optical quantum computing[25] require near-unity single-photon efficiency, and thus novel methods for optical control of single quantum emitters. One possibility is to use coherent two-photon excitation[26] to drive the cascaded biexciton-exciton transition in a quantum dot, where a record-high single-photon purity has been reported[27,28]. However, in such a three-level system, the generation of single-photon sources with simultaneous combination of single polarization, near-unity indistinguishability and efficiency was proven difficult.

Here, we propose and demonstrate a new way to coherently drive a quantum two-level system using a dichromatic field that has no spectral overlap with the transition frequency. As schematically shown in Fig. 1a, the dichromatic pulse consists of two sidebands, which have the same pulse envelope $\varepsilon(t)$, a fixed phase difference $\delta\varphi$ and are symmetrically red-detuned and blue-detuned by $\Delta\omega$ from the atomic transition frequency $\omega_{eg}$. A phase-coherent combination of the two sidebands, as shown by the mathematical relation,

$$\frac{\varepsilon(t)}{2}\cos((\omega_{eg}+\Delta\omega)\,t+\delta\varphi)+\frac{\varepsilon(t)}{2}\cos((\omega_{eg}-\Delta\omega)\,t)$$
$$=\varepsilon(t)\cos(\Delta\omega t+\frac{\delta\varphi}{2})\cos(\omega_{eg}\,t+\frac{\delta\varphi}{2}),$$

effectively cancels the detuning, and results into a resonant pulse with a new envelope $E(t)=\varepsilon(t)\cos(\Delta\omega t+\delta\varphi/2)$, as shown in Fig. 1b. The dichromatic pulse yields an effective Rabi frequency proportional to $E(t)$. The time-integrated pump pulse area can be suitably engineered to π, which will deterministically invert the population. Because the detuning can be set much larger than the linewidth of the resonance fluorescence, the single photons can be spectrally isolated from the well-separated laser background with a bandpass filter.

Assuming a Gaussian-shaped envelope, the time-resolved Rabi frequency is plotted in Fig. 1c for different ratios of the detuning ($\Delta\omega$) to the sideband linewidth ($\delta_{\mathrm{FWHM}}$), where a more rapid oscillation is observed for increasing $\Delta\omega/\delta_{\mathrm{FWHM}}$ ratios. Owing to the parity-alternating oscillations, the laser power to reach a π pulse is expected to increase for larger detunings. In addition, the relative phase between the red and blue sidebands gives rise to a new quantum interference phenomenon and a controlling knob for the two-level system beyond the conventional one-pulse resonant excitation[7,29]. As illustrated in Fig. 1d, the phase $\delta\varphi$ can control the relative position of the positive and the negative parts of the Rabi frequency within the envelope, thereby modulating the time-integrated pulse area.

The coherent dynamic process could be conceptually visualized as the motion of the well-known Bloch vector. While the red-detuned or blue-detuned sideband alone can't efficiently drive the two-level system along of the prime meridian of the Bloch sphere (Fig. 1e), a phase-coherent combination of the two sidebands together can achieve a deterministic popular inversion through an oscillating trajectory as shown in Fig. 1f.

Our scheme is in principle applicable in various scenarios involving a single two-level quantum system interacting with electromagnetic fields. Here, we report the first experimental demonstration. The two-level system is a single InGaAs quantum dot embedded inside a 2.5 μm-diameter micropillar, cooled to 3.6 K. The micropillar cavity features a relatively low quality factor of ~1,000 such that its cavity bandwidth can accommodate the spectral width of the dichromatic pulse. A confocal microscope is used for laser excitation of the quantum dot and collection of the emitted resonance fluorescence.

First, we prepare a phase-locked dichromatic pulse using a 4f-system, as shown in Fig. 2a. Picosecond pulses with a width of 3 ps from a Ti:sapphire laser are diffracted by the first grating, i.e., the different frequency components are distributed to different spatial directions. A block placed in the Fourier plane intercepts the central frequency of the pulses. The remaining red and blue sidebands are then recombined on the second grating, and collected into a single-mode fiber. The central wavelength of the pulses is adjusted so that the quantum dot emission line sits in the center of the two sidebands. The spectrum of the filtered laser pulse in our experiment is shown in Fig. 2b, with a splitting of 132 GHz between the red and blue sideband, which is 82 times larger than the linewidth (~1.6 GHz) of the quantum-dot single-photon source.

The collection arm employs a similar 4f-system as that used in the excitation arm, except that a slit replaces the block, which passes through only the central frequency part of the single-photon emission while intercepting the excitation laser sidebands. In this experiment, such a double 4f-system effectively suppresses the laser scattering by an extremely high extinction ratio of $10^{11}$ (Supplementary Fig. 1). The transmission efficiency of the second 4f-system is measured to be 26.4% which in the future can be replaced by commercially available ultra-narrowband optical filters with bandwidth as narrow as 0.1 nm and near-unity transmission rate.

Next, we send the dichromatic pulses to excite a single-electron-charged quantum dot with the center of the dichromatic pulse being tuned to be resonant with the atomic transition, and test its performance in population inversion. Figure 3a plots the pulsed resonance fluorescence single photon intensity as a function of the driving optical field strength. For a comparison, the data from the conventional resonant excitation using a 3-ps laser pulse and cross-polarization extinction is also presented in Fig. 3a. The dichromatic excitation reaches a full population inversion at a pump power of 115 nW, 8.5 times larger than the monochromatic excitation. The single-photon count under the dichromatic excitation, after correcting for the transmission loss of the second 4f-system, is 1.74 times higher than that using the monochromatic excitation and cross-polarization. Such an enhancement is owing to the removal of the polarization filtering that usually sacrifice ~50% of the single photons. Ideally, the efficiency enhancement should be a factor of 2. The additional loss might be due to the damping at high-power regime (115 nW) under the influence of a phonon environment[30].

An important distinction of the solid-state emitter from an ideal two-level system is the presence of phonon. Phonon-assisted far off-resonant optical excitations have been demonstrated in previous work[31,32] which requires high pump power (typically >100 times the π pulse power with resonant excitation). We perform controlled experiments by isolating the red or blue sideband only to pump the quantum dot. Figure 3b shows the single photon intensity as a function of pump power. With a laser power of ~115 nW, which corresponds to the π pulse of the two-color excitation, the isolated phonon-assisted blue and red sideband excitation, however, can only achieve 15% and 7% of population inversion, respectively, in agreement with the theoretical model[31]. This data reinforces our model in Fig. 1 that because the two-color pulses are phase-locked, they should be effectively viewed as a resonant single-color pulse.

Next, we send the two-color pulse through a stable Sagnac interferometer as shown in Fig. 4a, where the red and blue sidebands are split into two paths with an adjustable time delay (thus the phase $\delta\varphi$) and then recombined on the output beam splitter. Figure 4b shows the detected resonance fluorescence counts as a function of the delay for two examples of driving field strength. In the weak excitation regime (bottom panel), the population of the excited state is approximately proportional to the effective input pulse area which is modulated by the relative phase $\delta\varphi$, thus the oscillation of the resonance fluorescence counts with the pulse delay is close to a sinusoidal function. This phenomenon can also be described as the coherent superposition of the excited state wave function created by two pulses in the low-excitation limit[7,29]. In the strong excitation regime (upper panel), the photon counts are dependent on the effective pulse area. The photon counts will drop when the pulse area is larger than the $\pi$ pulse as the relative phase $\delta\varphi$ varies, so a dip will appear at the peak of the sinusoid-like oscillation. A 2D map of the phase-sensitive interference fringes at varying phase delay and excitation strength is shown in Fig. 4c, which is in good agreement with numerical simulations (see Supplementary Information) in Fig. 4d, showing a gradual transition from a sinusoidal fringe in the weak power regime to a more complex structure in the strong power regime. We note that the minimum value in Fig. 4b does not drop to zero, which is mainly due to the phonon-induced dephasings[30] and remaining asymmetry of the red and blue pulses (such as the amplitude, linewidth) in our experiment so that the effective pulse area starts from a non-zero value.

Finally, we study the effect of this new coherent control technology on the purity and the indistinguishability of the emitted single photons, two key parameters for optical quantum computation[25,33], boson sampling[24] and quantum networks[34]. We first test the purity by second-order coherence measurements. Under $\pi$-pulse excitation and at zero relative phase between the two sidebands, we observe a vanishing multiphoton probability of $g^2(0) = 0.012 \pm 0.001$ in the collected photons (Fig. 5a). The photon indistinguishability is tested by using Hong-Ou-Mandel interference between two consecutively emitted single photons at a time delay of 13 ns. Figure 5b plots the time-delayed histograms of normalized counts for two photons prepared at orthogonal and parallel polarization, where the counts for the latter is significantly suppressed at zero time delay. From Fig. 5b, a degree of indistinguishability of $0.964 \pm 0.006$ is obtained between the two π-pulse excited single photons. A closer inspection of the co-polarized two-photon counts around the zero delay shows a dip. This is due to temporal filtering by ultrafast timing resolution (~20 ps) of the superconducting nanowire single-photon detectors, which increase the interference visibility at the dip to $0.982 \pm 0.006$. These results indicate a high level of photon coherence using the dichromatic pulses excitation, with a similar quality to the purity and indistinguishability data obtained from the one-color resonant excitation[9,14-20] (see Supplementary Information). Finally, we note that the high degree of photon indistinguishability suggests that the dichromatic excitation induce no additional time jitter, thus further excluding the model of phonon-assisted excitation.

In summary, we have proposed and demonstrated coherent driving the two-level system using dichromatic pulses. We have experimentally confirmed its ability in complete population inversion and phase control of the excitonic two-level system. In addition, it allows the production of high-purity and highly indistinguishable single photons. Our work provides additional degree of freedom to separate the excitation laser and the single photon, and can be combined with other techniques such as side excitation[16,35] and polarized microcavities[36] to generate truly optimal single-photon sources.

**Online content**

Any methods, additional references, Nature Research reporting summaries, source data, statements of data availability and associated accession codes are available.

**Data availability statement**

The data that support the plots within this paper and other findings of this study are available from the corresponding author upon reasonable request.

**Competing financial interests**

The author declare that they have no competing financial interests.

**References:**

1. Rabi, I. I., Millman, S. and Kusch, P. The molecular beam resonance method for measuring nuclear magnetic moments. The magnetic moments of $_3Li^6$, $_3Li^7$ and $_9F^{19}$. *Phys. Rev.* **55**, 526 (1939).

2. Ramsey, N. F. A Molecular beam resonance method with separated oscillating fields. *Phys. Rev.* **78**, 695 (1950).

3. Autler, S. H. and Townes, C. H. Stark effect in rapidly varying fields. *Phys. Rev.* **100**, 703 (1955).

4. Mollow, B. R. Power spectrum of light scattered by two-level systems. *Phys. Rev.* **188**, 1969 (1969).

5. Bennett, C. H. and DiVincenzo, D. P. Quantum information and computation, *Nature* **404**, 247-255 (2000).

6. Leibfried, D., Blatt, R., Monroe, C. and Wineland, D. Quantum dynamics of single trapped ions. *Rev. Mod. Phys.* **75**, 281-324 (2003).

7. Bonadeo, N. H., Erland, J., Gammon, D., Park, D., Katzer, D. S., & Steel, D. G. Coherent optical control of the quantum state of a single quantum dot. *Science.* **282**, 1473-1476 (1998).

8. Press, D., Ladd, T. D., Zhang, B. and Yamamoto, Y. Complete quantum control of a single quantum dot spin using ultrafast optical pulses. *Nature*, **456**, 218-221 (2008).

9.  Lodahl, P., Mahmoodian, S., & Stobbe, S. Interfacing single photons and single quantum dots with photonic nanostructures. *Rev. Mod. Phys.* **87**, 347-400 (2015).

10. Weber, J. R. *et al.* Quantum computing with defects. *Proc. Natl. Acad. Sci. USA* **107**, 8513–8518 (2010).

11. Clarke, J. and Wilhelm, F. K. Superconducting quantum bits. *Nature.* **453**, 1031-1042 (2008).

12. Zrenner, A., Beham, E., Stufler, S., Findeis, F., Bichler, M. and Abstreiter, G. Coherent properties of a two-level system based on a quantum-dot photodiode. *Nature*, **418**, 612-214 (2002).

13. Stievater, T. H. *et al.* Rabi oscillations of excitons in single quantum dots. *Phys. Rev. Lett.* **87**, 133603 (2001).

14. Michler, P. *et al.* A. A quantum dot single-photon turnstile device. *Science.* **290**, 2282-2285 (2000).

15. Santori, C., Fattal, D., Vučković, J., Solomon, G. S. & Yamamoto, Y. Indistinguishable photons from a single-photon device. *Nature (London)* **419,** 594-597 (2002).

16. Muller, A. *et al.* Resonance fluorescence from a coherently driven semiconductor quantum dot in a cavity. *Phys. Rev. Lett.* **99**, 187402 (2007).

17. Vamivakas, A. N., Zhao, Y., Lu, C.-Y. & Atature, M. Spin-resolved quantum-dot resonance fluorescence. *Nature Phys.* **5**, 198-202 (2009).

18. He, Y.-M. *et al.* On-demand semiconductor single-photon source with near-unity indistinguishability. *Nature Nanotech.* **8**, 213-217 (2013).

19. Buckley, S., Rivoire, K. & Vučković, J. Engineered quantum dot single-photon sources. *Rep. Prog. Phys.* **75**, 126503 (2012).

20. Senellart, P., Solomon, G. & White, A. High-performance semiconductor quantum-dot single-photon sources. *Nature Nanotech.* **12**, 1026-1039 (2017).

21. Ding, X. *et al.* On-Demand single photons with high extraction efficiency and near-unity indistinguishability from a resonantly driven quantum dot in a micropillar. *Phys. Rev. Lett.* **116**, 020401 (2016).

22. Somaschi, N. *et al.* Near-optimal single-photon sources in the solid state. *Nature Photon.* **10**, 340-345 (2016).

23. Wang, H. *et al.* Near-transform-limited single photons from an efficient solid-state quantum emitter. *Phys. Rev. Lett.* **116**, 213601 (2016).

24. Aaronson, S. & Arkhipov, A. The computational complexity of linear optics. In Proceedings of the ACM Symposium on Theory of Computing (ACM, New York, 2011), pp. 333-243.

25. Kok, P., Munro, W. J., Nemoto, K., Ralph, T. C., Dowling, J. P. & Milburn, G. J. Linear optical quantum computing with photonic qubits. *Rev. Mod. Phys.* **79**, 135-174 (2007).

26. Müller, Markus, et al. On-demand generation of indistinguishable polarization-entangled photon pairs. Nature Photonics **8**, 224 (2014).

27. Schweickert, Lucas, et al. On-demand generation of background-free single photons from a solid-state source. *Applied Physics Letters* **112**, 093106 (2018).

28. Hanschke, Lukas, et al. Quantum dot single-photon sources with ultra-low multi-photon probability. *npj Quantum Information* **4**, 43 (2018).

29. Htoon, H. *et al.* Interplay of Rabi oscillations and quantum interference in semiconductor quantum dots. *Phys. Rev. Lett.* **88**, 087401 (2002).

30. Förstner, J., et al. Phonon-assisted damping of Rabi oscillations in semiconductor quantum dots. *Phys. Rev. Lett.* **91**, 127401 (2003).

31. Glässl, M., A. M. Barth, and V. M. Axt. Proposed robust and high-fidelity preparation of excitons and biexcitons in semiconductor quantum dots making active use of phonons. *Phys. Rev. Lett.* **110**, 147401 (2013).

32. Quilter, J. H., et al. Phonon-assisted population inversion of a single InGaAs/GaAs quantum dot by pulsed laser excitation. *Phys. Rev. Lett.* **114**, 137401 (2015).

33. Pan, J. W., Chen, Z. B., Lu, C.-Y., Weinfurter, H., Zeilinger, A., & Żukowski, M.

Multiphoton entanglement and interferometry. *Rev. Mod. Phys*. **84**, 777-838 (2012).

34. Humphreys, P. C. Deterministic delivery of remote entanglement on a quantum network. *Nature* **558**, 268-273 (2018).

35. Ates, S., Ulrich, S. M., Reitzenstein, S., Löffler, A., Forchel, A., & Michler, P. Post-selected indistinguishable photons from the resonance fluorescence of a single quantum dot in a microcavity. *Phys. Rev. Lett.* **103**, 167402 (2009).

36. He, Y.-M. *et al.* Polarized indistinguishable single photons from a quantum dot in an elliptical micropillar. *arXiv*:1809.10992 (2018).

**Figure captions:**

**Figure 1 | Coherent driving of a single two-level system with dichromatic pulses.**
**a.** Upper panel: energy level structure of a two-level system and energy separation of two detuned optical pulses (red and blue arrows); bottom panel: the dichromatic pulse in the frequency domain. It combines two phase-locked red- and blue-detuned pulses, symmetrically detuned by $\Delta\omega$ from the atomic transition $\omega_{eg}$. **b.** The dichromatic pulses in the time domain. **c.** Orange solid line: the time-resolved Rabi frequency for different detuning $\Delta\omega$, where the relative phase $\delta\varphi$ is fixed to 0; blue dashed line: the Gaussian envelopes $\varepsilon(t)$ of the red- (blue-) detuned pulse with a full width at half maximum of $\delta_{\text{FWHM}}$. When $\Delta\omega/\delta_{\text{FWHM}}$ increases from 2 to 4, the Rabi frequency oscillates faster between positive and negative parts. **d.** Orange solid line: the time-resolved Rabi frequency for different relative phases ( $\delta\varphi = 0, \pi, 2\pi$ ) for fixed $\Delta\omega = 2\delta_{\text{FWHM}}$. **e.** The trajectory of a qubit on the Bloch sphere, driven by a far red- or blue- detuned pulse alone. **f.** The trajectory of a qubit driven by the dichromatic pulse.

**Figure 2 | Generation of dichromatic pulses using a 4f optical system. a.** A 3-ps

pulse is diffracted by a grating such that the different frequency components focus at different location at the Fourier plane. A block placed at the center of the optical beam to utilized to cut out the central frequency component. A second grating is applied at the symmetric position to recover the spatial mode of the optical beam. **b**. The spectra of the dichromatic pulses created in our experiment and the quantum dot emission line. The spectra are shifted slightly vertically for clarity.

**Figure 3 | Single photon intensity as a function of the driving strength. a.** The Rabi oscillations under one-color and two-color excitations. Under two-color excitation, the single-photon intensity oscillates near sinusoidally as the pump power increase (purple circle dots). Compared to one-color excitation (open square dots), the maximum rate is 1.74 higher after corrected the transmission efficiency of the 4f system, while the excitation power is 8.5 times higher. See Supplementary Information for the detailed budge of the optical elements. **b.** The single photon intensity as a function of the driving power under dichromatic excitation (purple circle dots), isolated phonon-assisted blue (blue dots) and red (red dots) sideband excitation. As increasing the driving power, the counts climb slowly under the blue and red sideband excitation. When the two-color excitation reaches the population inversion under the driving power of ~115 nW, the blue and red sideband excitation can only achieve 15% and 7% of the popular inversion, respectively.

**Figure 4 | Phase-dependent resonance fluorescence under dichromatic driving. a**. A Sagnac interferometer was employed to stably control the phase of the dichromatic laser pulses. First, two filters at two different paths filter the red- and blue-detuned pulse, respectively. Then, one phase shifter was inserted into one of the paths to adjust the relative phase. **b**. Under low excitation (bottom panel), the quantum interference of probability waves excited by the red- and blue-shifted pulse gives rise to sinusoidal oscillation of the photon counts when phase changes. Under strong excitation (top

panel), the photon counts depend on the effective pulse area. When the pulse area is larger than π, there will be a dip at the peak of the sinusoidal oscillation. **c.** Phase-dependent quantum interference and Rabi oscillations under different excitation power. **d.** Theoretical simulation of the phase-dependent quantum interference and Rabi oscillations.

**Figure 5 | Characterization of the single-photon source under two-color excitation.**
**a.** Intensity-correlation histogram which shows a second-order correlation function of $g^2(0) = 0.012 \pm 0.001$. **b.** The extracted photon indistinguishability is $0.964 \pm 0.006$ by integrating the whole peak. By considering only the central dip resolved by fast detection, the interference visibility is improved to $0.982 \pm 0.006$.

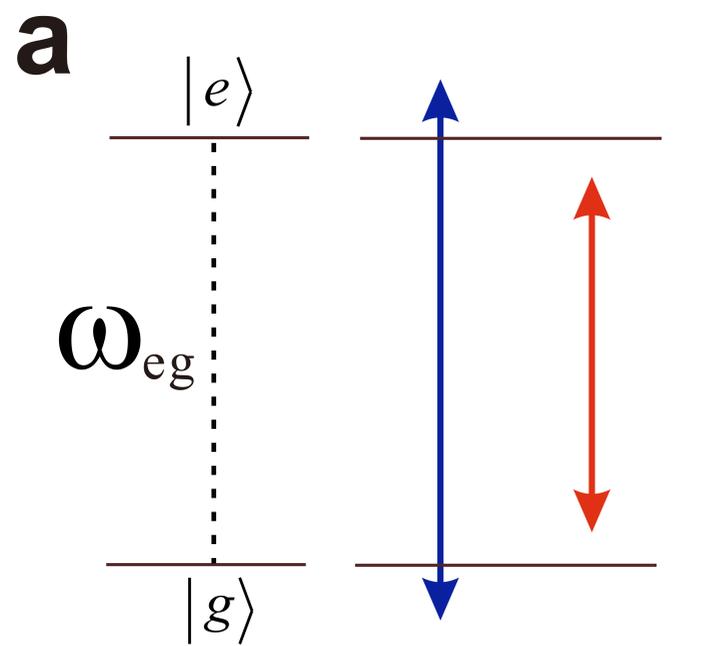

**a**

$|e\rangle$

$\omega_{eg}$

$|g\rangle$

phase-locked

$-\Delta\omega \quad \omega_{eg} \quad +\Delta\omega \quad \omega$

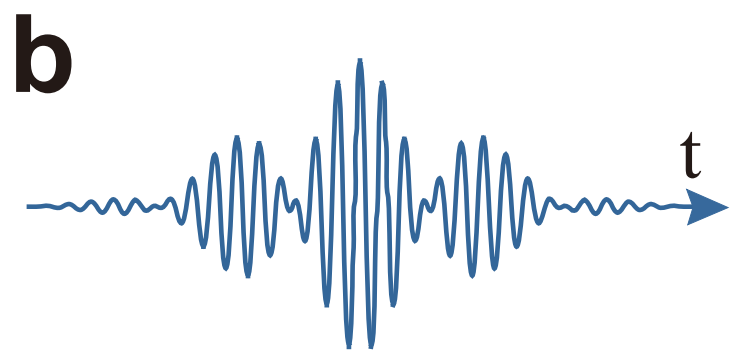

**b**

$t$

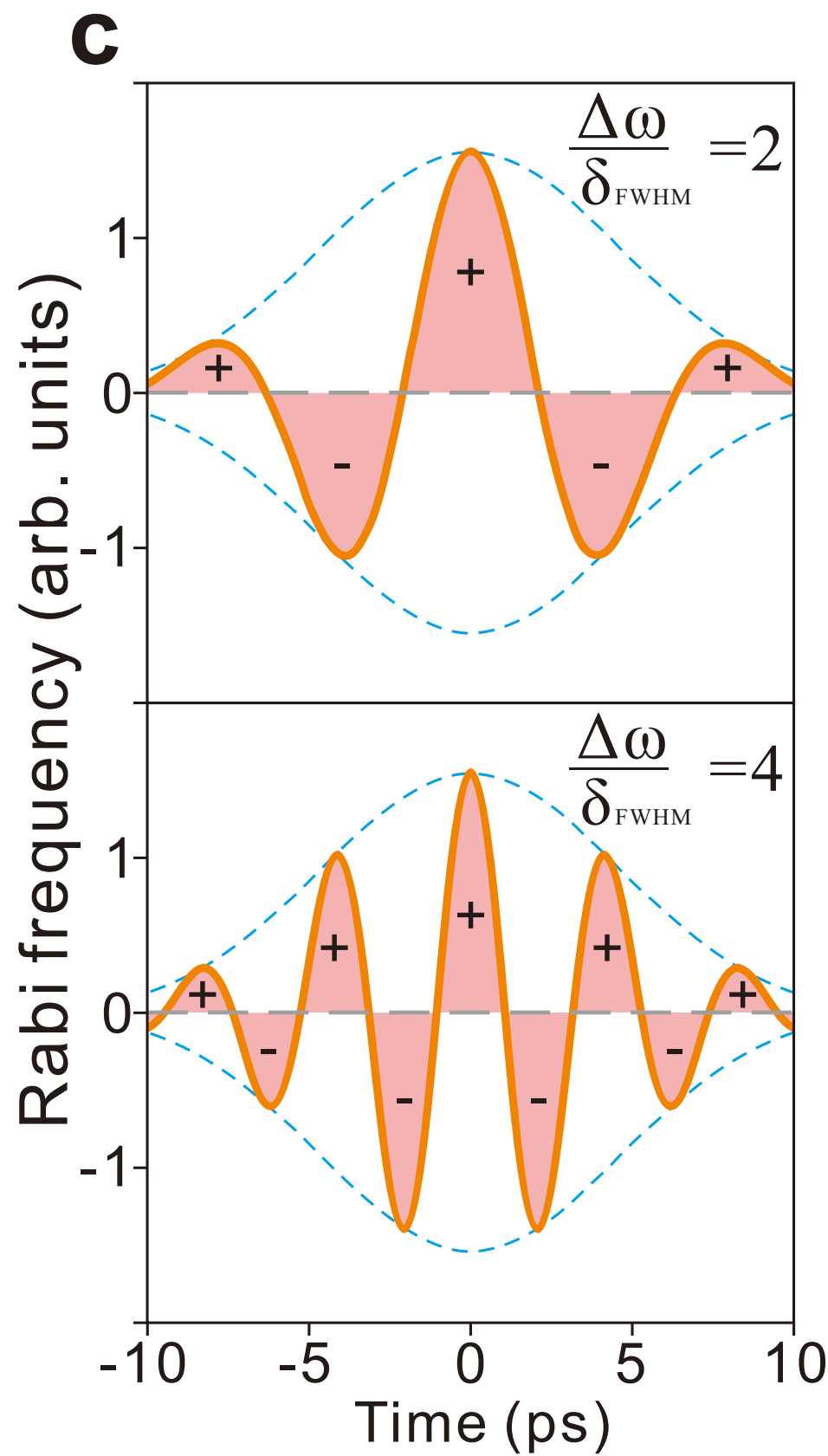

**c**

$\frac{\Delta\omega}{\delta_{FWHM}}=2$

$\frac{\Delta\omega}{\delta_{FWHM}}=4$

Rabi frequency (arb. units)

Time (ps)

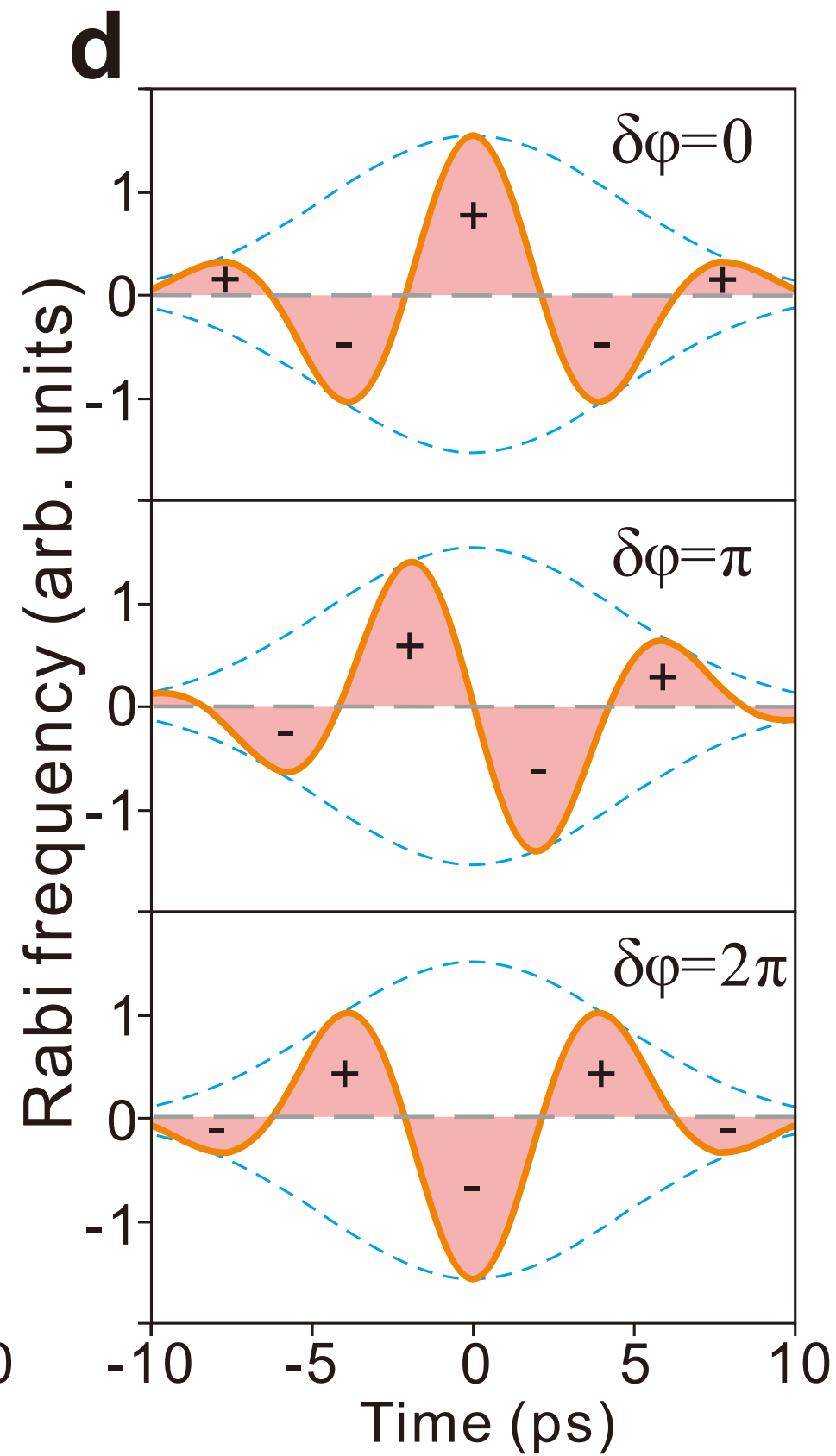

**d**

$\delta\varphi=0$

$\delta\varphi=\pi$

$\delta\varphi=2\pi$

Time (ps)

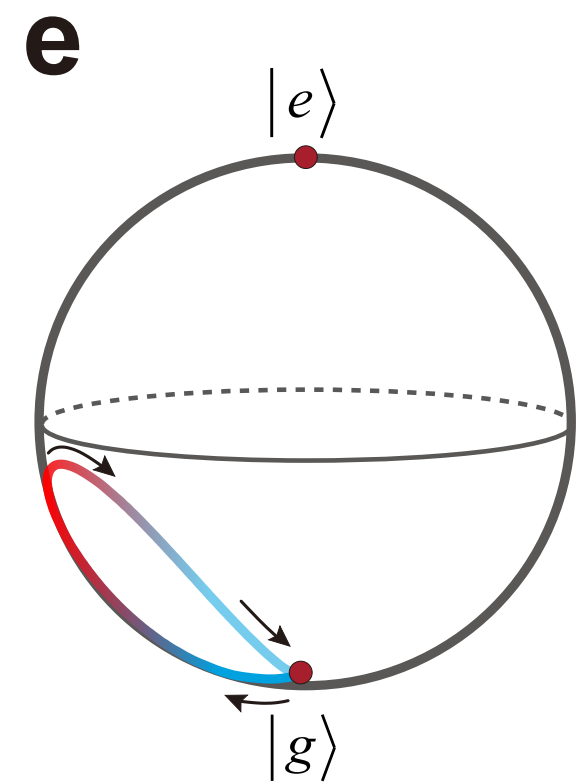

**e**

$|e\rangle$

$|g\rangle$

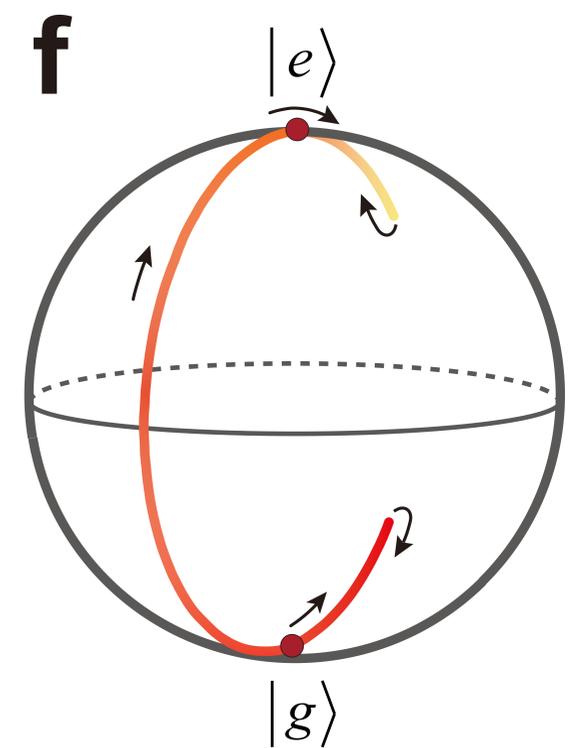

**f**

$|e\rangle$

$|g\rangle$

**a**

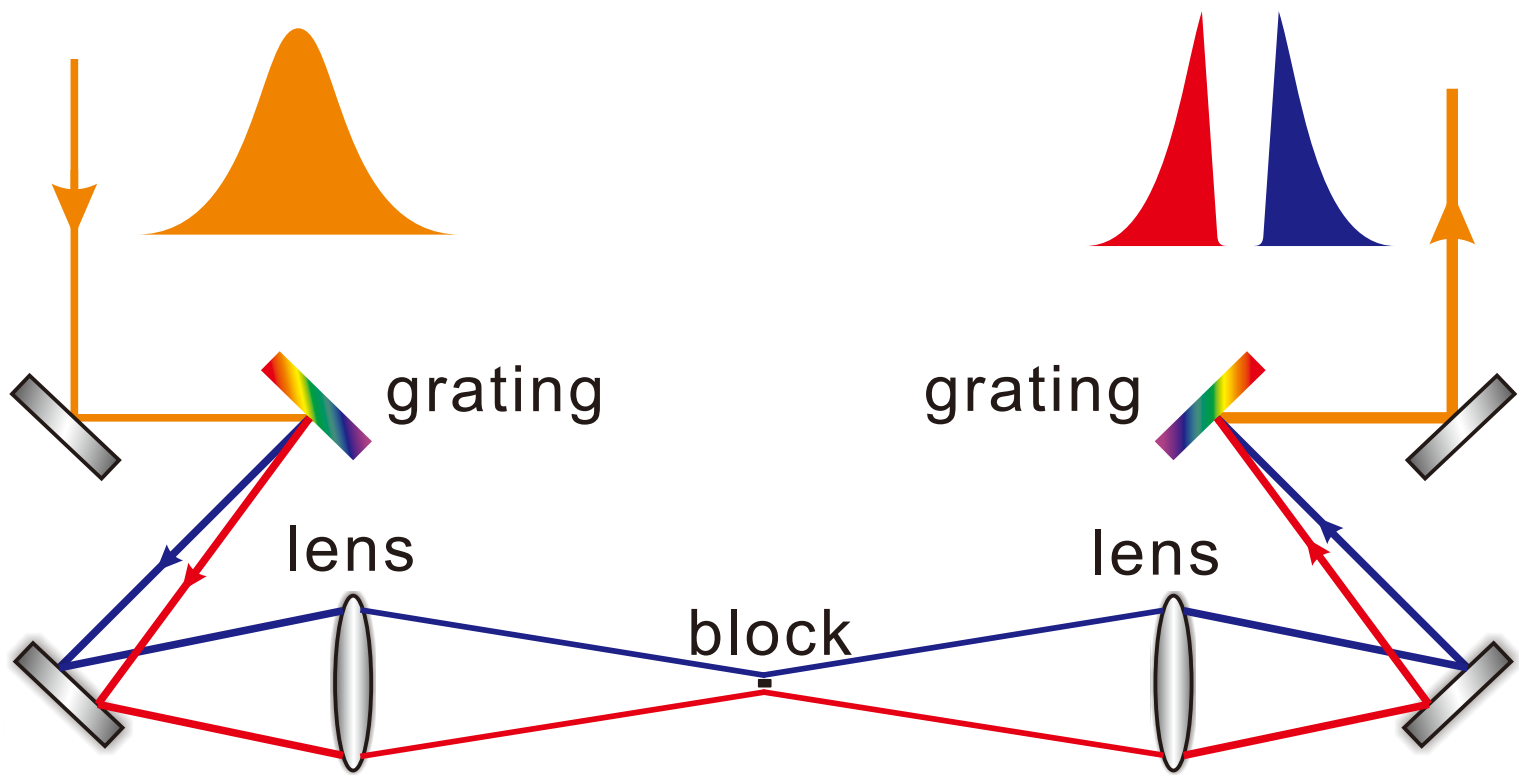

grating

lens

block

lens

grating

**b**

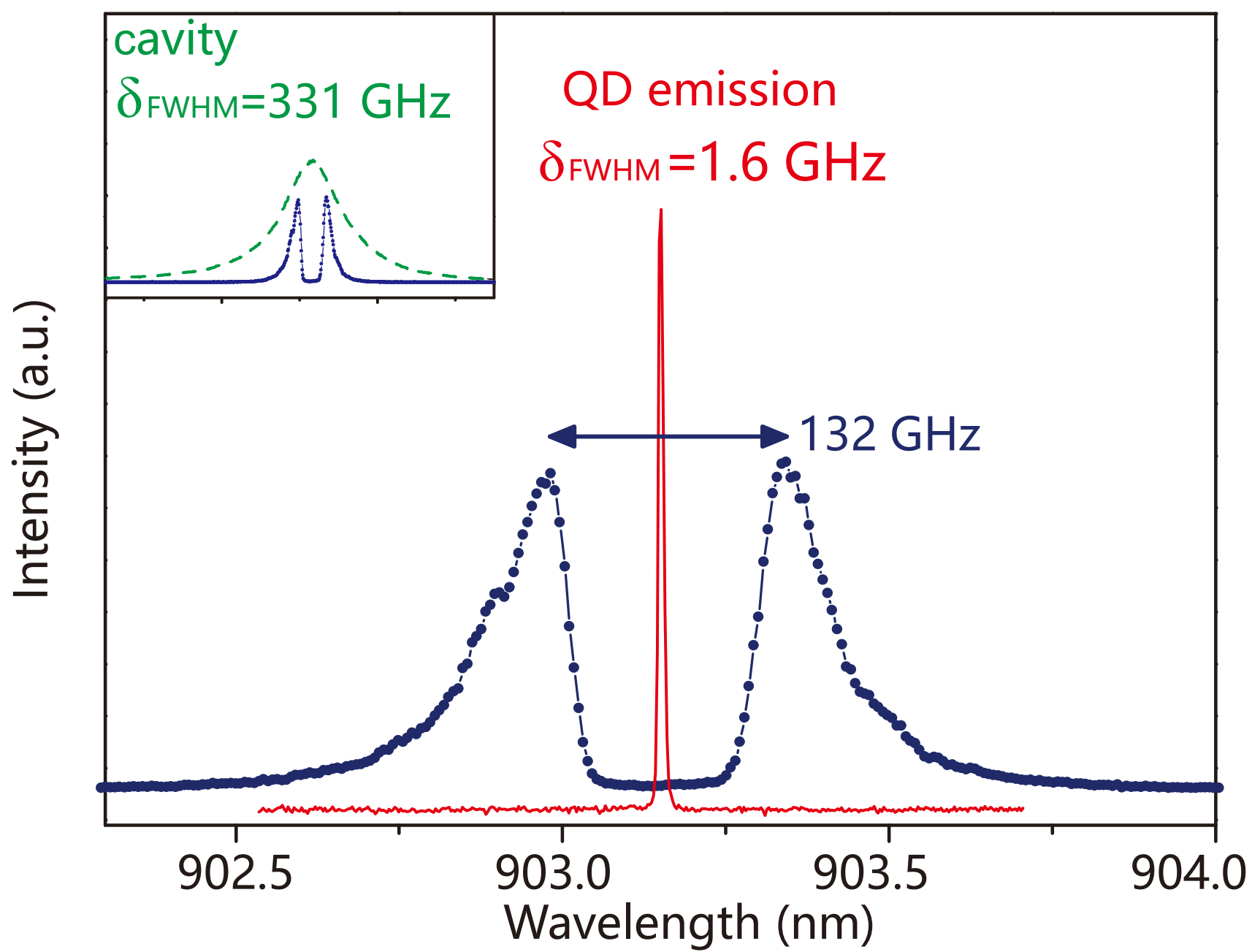

cavity
$\delta_{FWHM}=331$ GHz

QD emission
$\delta_{FWHM}=1.6$ GHz

132 GHz

Intensity (a.u.)

Wavelength (nm)

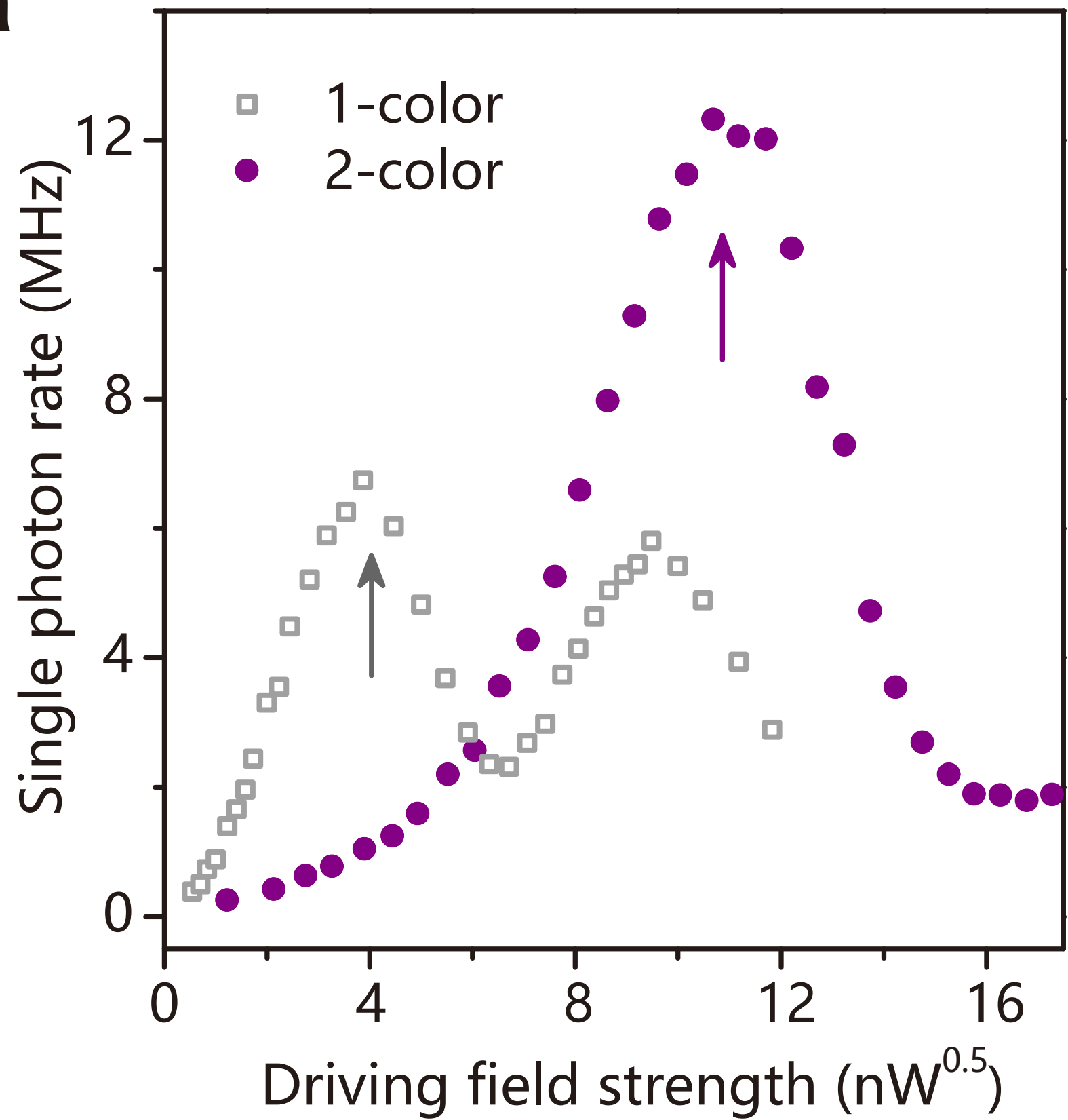

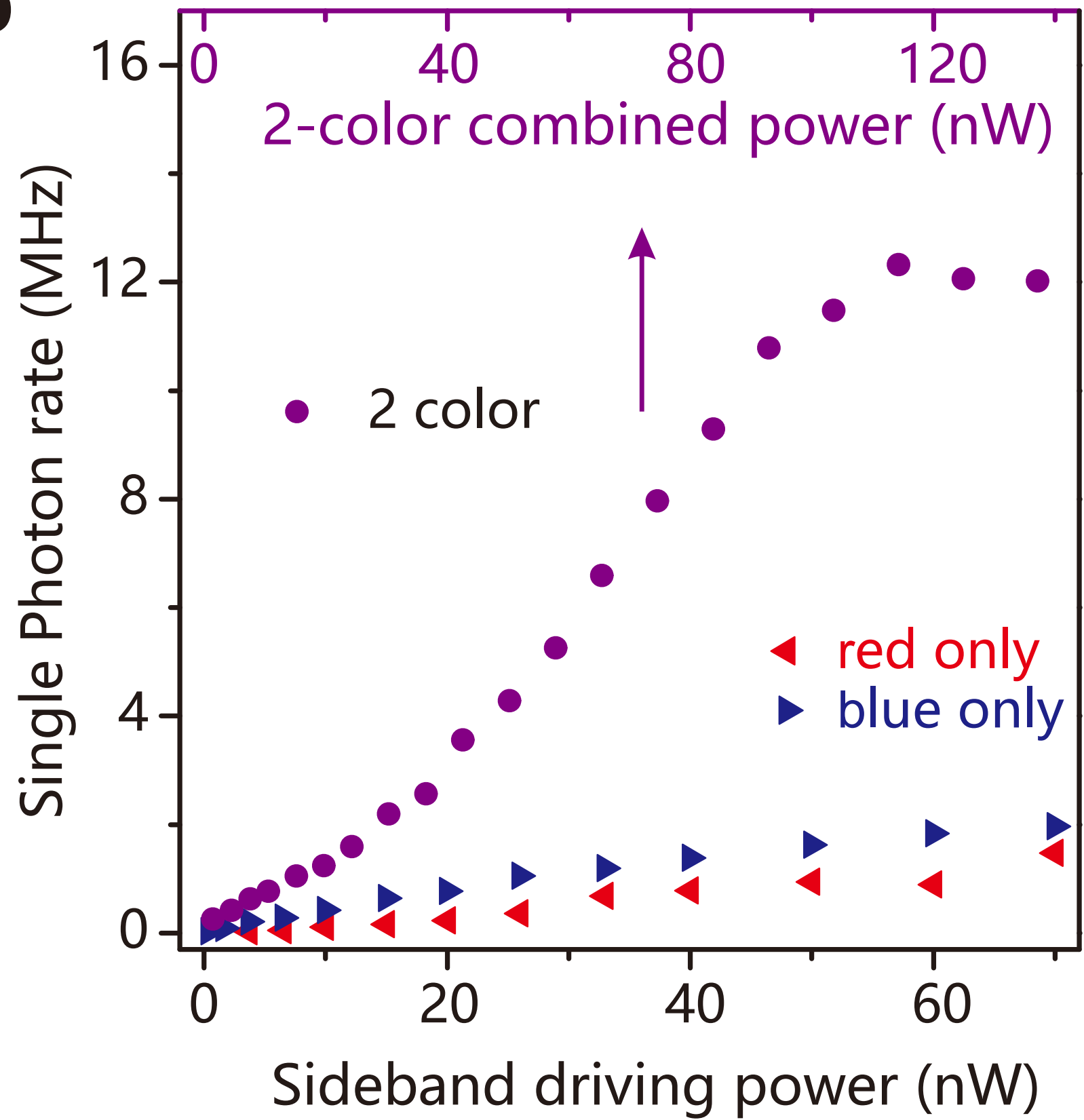

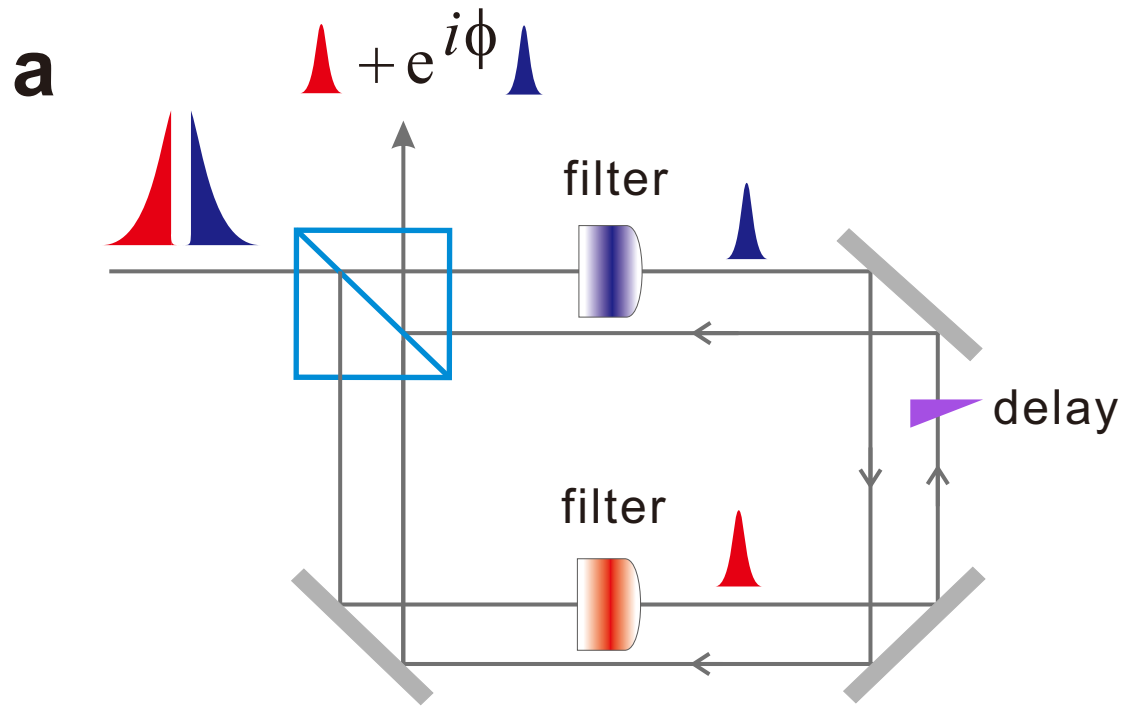

**a**

$\textcolor{red}{\blacktriangle} + e^{i\phi} \textcolor{blue}{\blacktriangle}$

filter

delay

filter

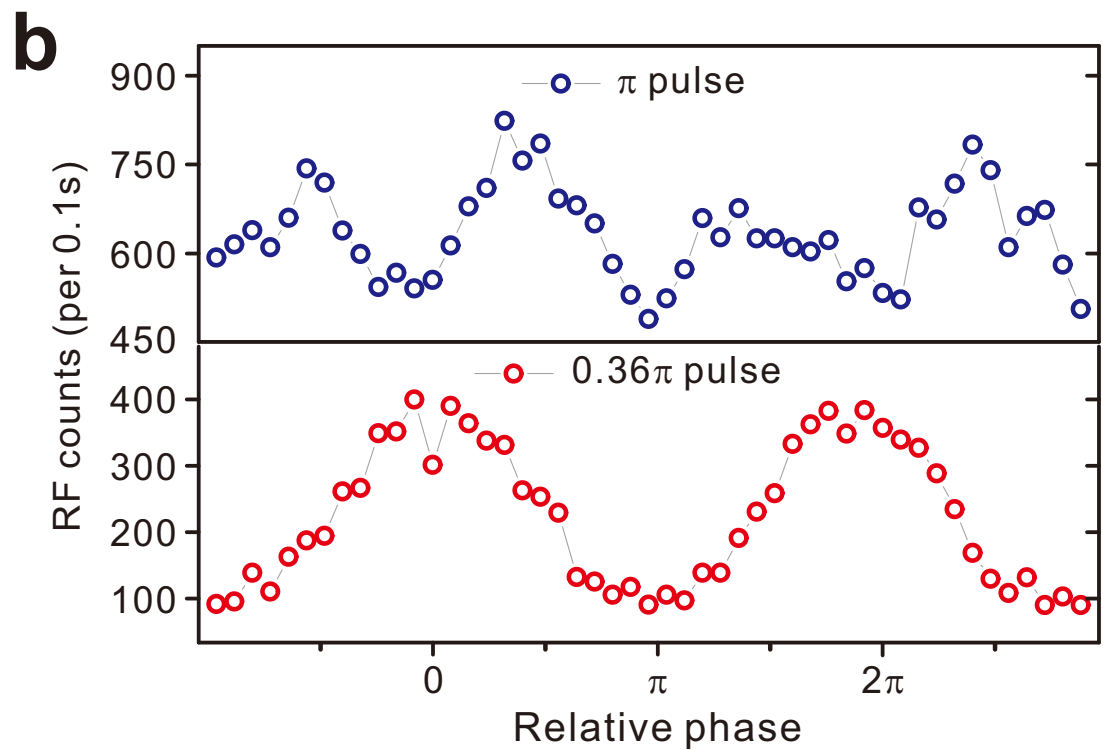

**b**

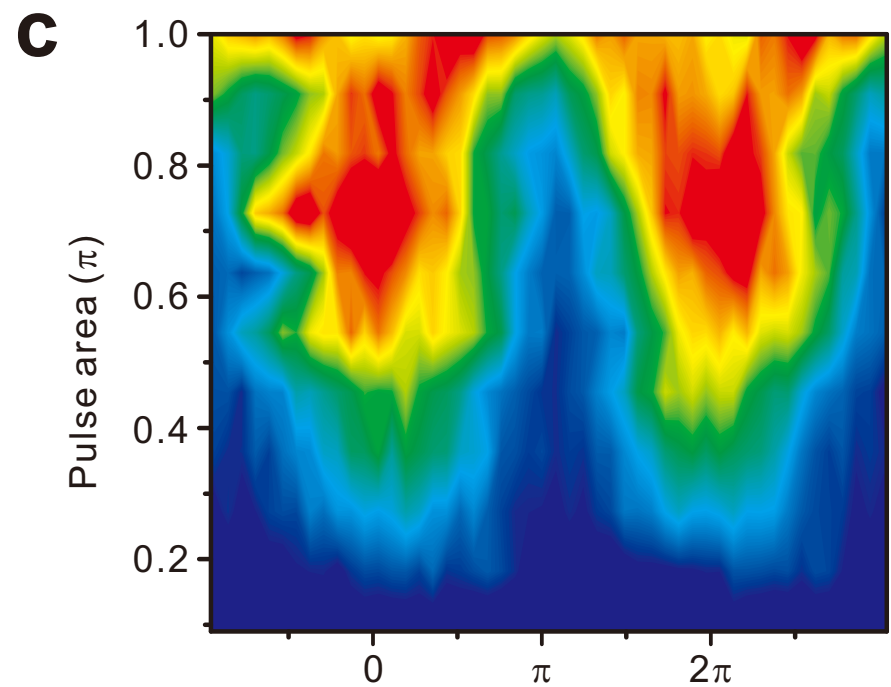

**c**

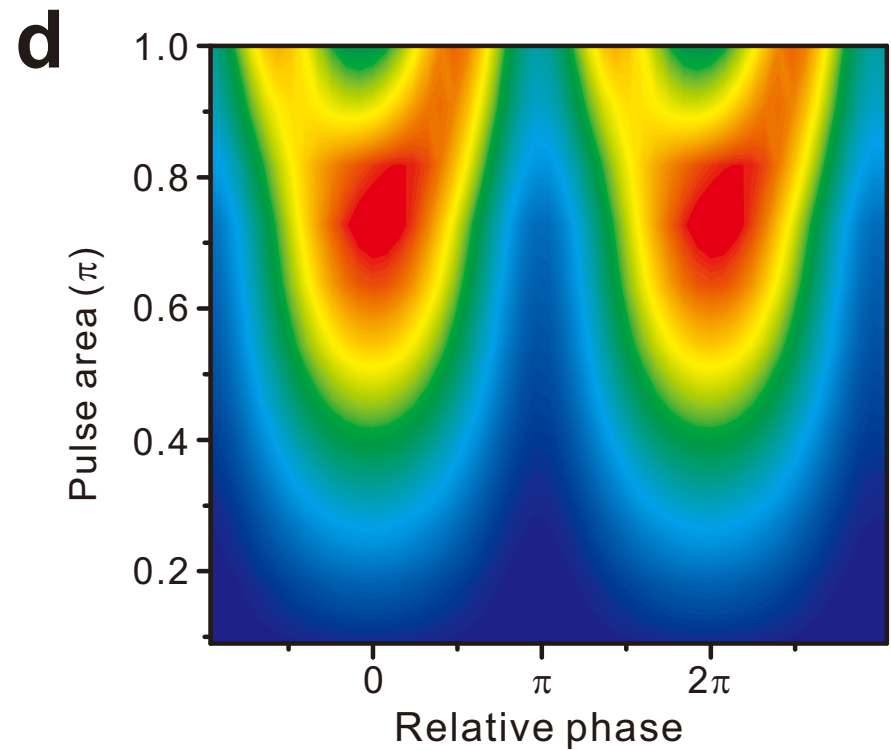

**d**

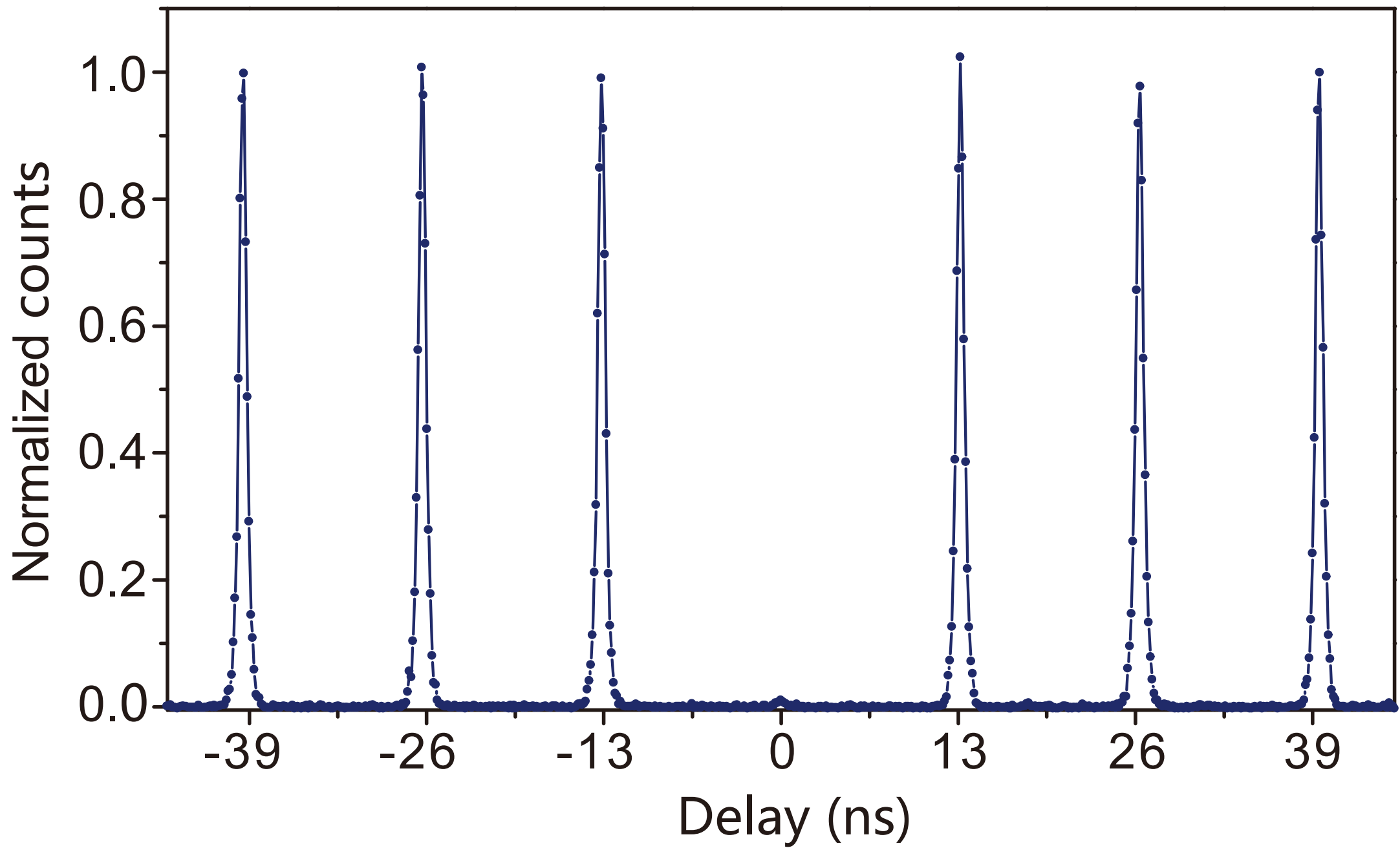

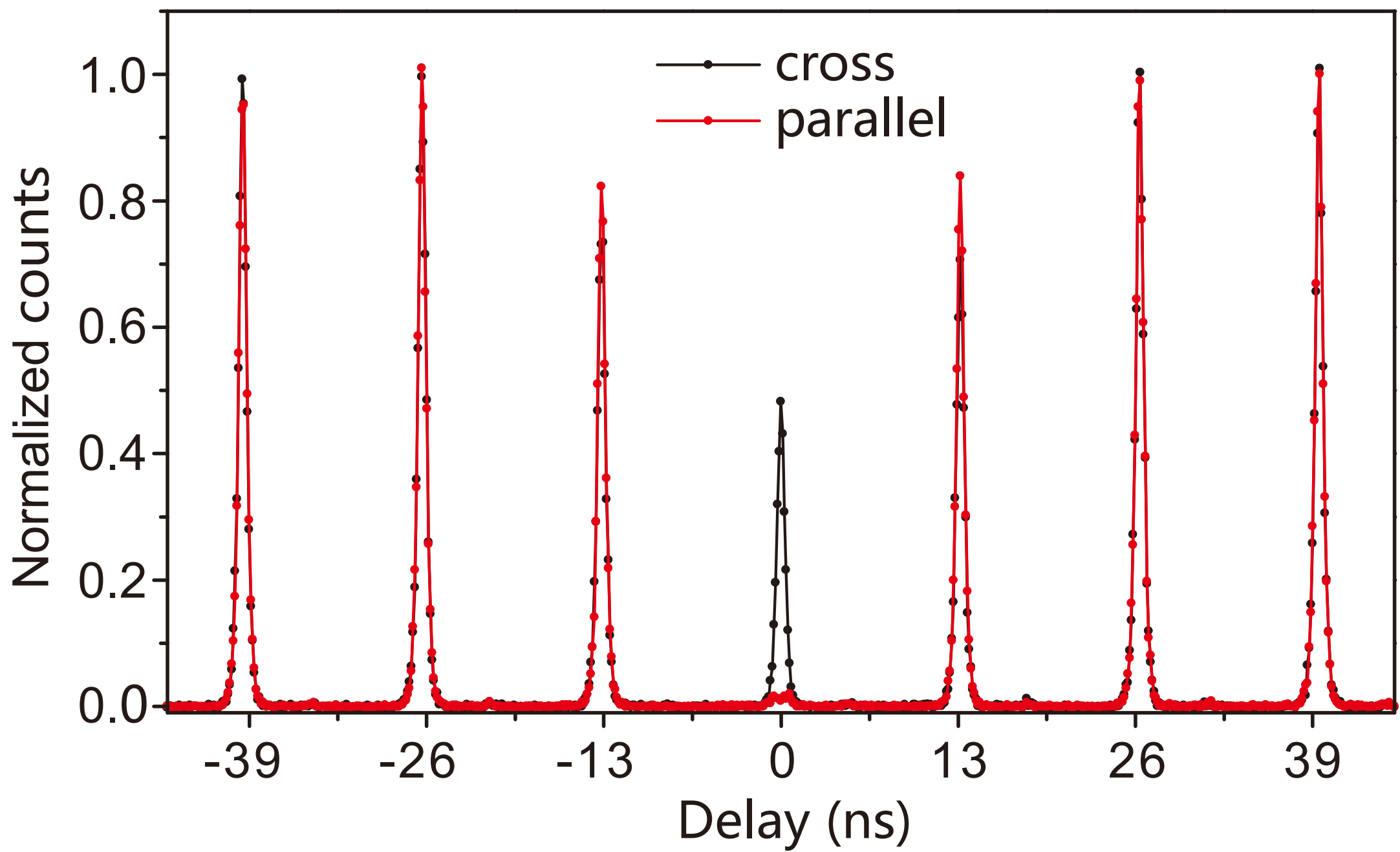